\documentclass[journal=jctcce,manuscript=article]{achemso}

\usepackage[version=3]{mhchem} 
\usepackage[T1]{fontenc}
\usepackage{tipa}
\usepackage{hyperref}

\usepackage{amssymb, amsmath}
\usepackage{braket}
\usepackage{bm}
\usepackage{upgreek}
\newcommand{\mr}{\mathrm}

\newcommand*{\matr}[1]{\mathbf{#1}}
\newcommand*{\vect}[1]{\bm{#1}}

\usepackage[authormarkuptext=name,authormarkup=none,final]{changes}
\definecolor{red}{RGB}{255, 0, 0}
\definecolor{blue}{RGB}{0, 0, 255}
\definecolor{green}{RGB}{0, 192, 0}
\definechangesauthor[name={Gianluca}, color={red}]{GL}
\definechangesauthor[name={Yorick}, color={blue}]{YLAS}
\definechangesauthor[name={Alec}, color={green}]{AES}

\usepackage{tablefootnote}
\usepackage{fancyhdr}

  \newcommand*{\citen}{}
\DeclareRobustCommand*{\citen}[1]{%
  \begingroup
    \romannumeral-`\x 
    \setcitestyle{numbers}%
    \cite{#1}%
  \endgroup
}

\author{Elli Selenius}
\affiliation{Science Institute of the University of Iceland, Reykjavík, Iceland}
\email{elliselenius@hi.is}
\author{Alec Elías Sigurdarson}
\affiliation{Science Institute of the University of Iceland, Reykjavík, Iceland}
\author{Yorick L. A. Schmerwitz}
\affiliation{Science Institute of the University of Iceland, Reykjavík, Iceland}
\author{Gianluca Levi}
\affiliation{Science Institute of the University of Iceland, Reykjavík, Iceland}
\email{giale@hi.is}

\title{Orbital-optimized versus time-dependent density functional calculations of intramolecular charge transfer excited states}

\begin{document}

\begin{tocentry}

    \includegraphics[width=\textwidth]{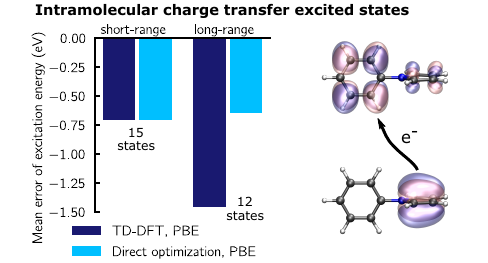}




\end{tocentry}

\begin{abstract}
The performance of time-independent, orbital optimized calculations of excited states is assessed with respect to charge transfer excitations in organic molecules in comparison to the linear-response time-dependent density functional theory (TD-DFT) approach. A direct optimization method to converge on saddle points of the electronic energy surface is used to carry out calculations with the local density approximation (LDA) and the generalized gradient approximation (GGA) functionals PBE and BLYP for a set of 27 excitations in 15 molecules. The time-independent approach is fully variational and provides a relaxed excited state electron density from which the extent of charge transfer is quantified. The TD-DFT calculations are generally found to provide larger charge transfer distances compared to the orbital optimized calculations, even when including orbital relaxation effects with the Z-vector method. While the error on the excitation energy relative to theoretical best estimates is found to increase with the extent of charge transfer up to ca.\ $-2$ eV for TD-DFT, no correlation is observed for the orbital optimized approach. The orbital optimized calculations with the LDA and the GGA functionals provide a mean absolute error of $\sim$0.7 eV, outperforming TD-DFT with both local and global hybrid functionals for excitations with long-range charge transfer character. Orbital optimized calculations with the global hybrid functional B3LYP and the range-separated hybrid functional CAM-B3LYP on a selection of states with short- and long-range charge transfer indicate that inclusion of exact exchange has a small effect on the charge transfer distance, while it significantly improves the excitation energy, with the best performing functional CAM-B3LYP providing an absolute error typically around 0.15 eV.  
\end{abstract}


\section{Introduction}
Excited electronic states of organic molecules can involve intramolecular charge transfer, where electronic charge is transferred from one fragment (donor) to another (acceptor) upon excitation from the ground state. Photoinduced intramolecular charge transfer plays an important role in natural processes, such as photosynthesis\cite{Zigmantas2002,Kosumi2012} and vision \cite{Tomasello2009}, and in various applications, such as dye-sensitised solar cells \cite{Mariotti2020}, organic light-emitting diodes \cite{Zou2020,Liu2018}, and simultaneous cancer therapy and imaging \cite{Liu2019b, Feng2016}. Modelling charge transfer excited states of organic chromophores is essential for gaining insights into these applications. Often, this involves simulating the dynamics of atoms on the excited state potential energy surfaces, requiring methods that provide reasonable accuracy at affordable computational cost. Highly accurate multireference configuration interaction and coupled cluster methods are employed for the calculation of reference energy values for small molecules and dimers \cite{Loos2021, Kozma2020}. However, these wave function methods become impractical for larger molecules relevant for applications because the computational cost increases rapidly with the system size. Furthermore, analytic atomic forces necessary for excited state molecular dynamics simulations are often not available.

Presently, the method of choice for calculations of excited states of large systems is the linear-response time-dependent density functional theory (TD-DFT) approach using the adiabatic approximation\cite{Casida1995, Runge1984, Hohenberg1964}, because of its relatively low computational cost. This approach is based on the use of ground state orbitals within linear response to describe the excitations. As a result, TD-DFT can achieve good results for valence excited states of molecules, but tends to fail for excitations involving significant changes in the electron density, where the ground state orbitals are not optimal for the excited state. Long-range charge transfer excitations represent one example, as charge transfer induces a large rearrangement of the electron density. In this case, the performance of TD-DFT is known to be highly dependent on the Kohn-Sham (KS)\cite{Kohn1965} exchange-correlation functional used. Computationally efficient local and semi-local functionals, but also more expensive global hybrid functionals, are found to underestimate the excitation energy\cite{wang2023earth, Mester2022, Hait2021, Zhao2019, Dev2012, Dreuw2004, Dreuw2003} and fail to give the $1/R$ dependence of the energy on the separation between donor and acceptor\cite{Dreuw2003}. Improved results are obtained with more advanced functionals, such as range-separated hybrids\cite{Shee2020, Kronik2012}, involving the addition of non-local exact exchange to the long-range part of the electron-electron interaction. However, the computational cost is increased and the calculations are complicated by the fact that the  optimal range separation is system-specific\cite{Korzdorfer2014}. As a result, the functional that can best describe a long-range charge transfer state is not always the best option for a more locally excited state \cite{Liang2022, Dev2012}, even for the same molecule \cite{Liu2019}. 

The search for a low-cost method that allows high-energy and charge transfer excitations to be accurately described has recently sparked renewed interest in variational, time-independent density functional calculations. In this conceptually simple approach, the orbitals are variationally optimized to find solutions to the KS equations higher in energy than the ground state. These mean-field solutions corresponding to single Slater determinants with non-aufbau occupation represent the excited states.\cite{Giarrusso2023, Hait2021, Ayers2015, Ayers2012, Gorling1999, Perdew1985} Thus, unlike in TD-DFT, both ground and excited states are treated variationally, resulting in a more balanced description of the different electronic states. This approach is sometimes called delta self-consistent field ($\Delta$SCF)\cite{Mazzeo2023, Vandaele2022, Kowalczyk2011}, referring to the calculation of the excitation energy as the energy difference between the individually optimized ground and excited state solutions. Compared to TD-DFT, orbital optimized density functional calculations have been shown to produce better results for long-range intermolecular charge transfer excitations\cite{Hait2021, Zhao2019, Barca2018, Briggs2015, Dreuw2004}. Firstly, they provide the correct $1/R$ asymptote of the energy, and secondly, they typically yield a more accurate excitation energy, even when local and semi-local functionals are used. Fewer studies address intramolecular charge transfer states where the separation between donor and acceptor is smaller, with the preliminary results indicating that the time-independent approach can also give an improvement in these cases\cite{Kumar2022, Hait2016, Zhekova2014, Himmetoglu2012}. 

The variational orbital optimization provides a relaxed excited state electron density, so state-specific implicit\cite{Kunze2021} or explicit\cite{Nottoli2022} solvation effects, which are important for the description of charge transfer excitations, can be included in the same way as for ground state calculations. Analytic atomic forces are also readily available via the Hellman-Feynman theorem to perform excited state molecular dynamics simulations, where the effect of the environment can be included through quantum mechanics/molecular mechanics (QM/MM) embedding\cite{Mazzeo2023, Vandaele2022, Malis2021, Levi2020pccp, Malis2020, Levi2018, Pradhan2018}. For example, a recent QM/MM study elucidated the structural dynamics following photoexcitation of a metal-to-ligand charge transfer state in a solvated copper complex photosensitizer\cite{Katayama2023, Levi2018}, while another modelled photoinduced proton-coupled electron transfer in a chromophore embedded in a photoreceptor protein\cite{Mazzeo2023}. 

One reason for the relatively limited number of studies on intramolecular charge transfer excitations is that the orbital optimized calculations face several practical challenges, which make convergence difficult. This is exemplified by the molecular dynamics work of ref \citen{Mazzeo2023}, where 10 out of 30 excited state trajectory propagations are reported to have failed due to SCF convergence issues. The main challenge is that excited state solutions are typically saddle points on the surface defined by the variation of the energy as a function of the electronic degrees of freedom\cite{Schmerwitz2023, Burton2022, Hait2021, Levi2020b, Perdew1985}. As a consequence, the calculations are prone to collapsing to lower-energy solutions along the degrees of freedom where the energy should be maximized, i.e.\ the instabilities of the target saddle point on the electronic energy surface. Several methods have been proposed to prevent this variational collapse. One of the most used is the maximum overlap method (MOM), where at each iteration of the optimization, the orbital occupations are chosen to maximize the overlap with a set of reference orbitals\cite{Gilbert2008, Barca2018}, typically the initial guess orbitals.\cite{Barca2018} However, for charge transfer excitations in organic molecules, MOM does not eliminate the risk of variational collapse completely, as shown recently for intramolecular charge transfer states of nitrobenzene and twisted $N$-Phenylpyrrole (PP)\cite{Schmerwitz2023, Mewes2014}. Preliminary studies show that the calculations can collapse to lower-energy solutions where the charge is too delocalized, giving an inadequate description of the excited state\cite{Schmerwitz2023, Hait2021}. Moreover, even when MOM manages to prevent variational collapse, the convergence can still be problematic when using SCF algorithms based on eigendecomposition of the Hamiltonian matrix, such as the direct inversion in the iterative subspace (DIIS)\cite{Ivanov2021, Obermeyer2021, Carter2020, Hait2020, Levi2020b, Levi2020a}. 

Recently, direct orbital optimization (DO) methods have been developed for time-independent, variational excited state calculations\cite{Schmerwitz2023, Ivanov2021, Levi2020b, Levi2020a}. These approaches use approximate second-order optimization algorithms for locating saddle points akin to those for transition state searches in atomic rearrangements, often in combination with MOM (DO-MOM). DO methods have proven more robust than DIIS-type alternatives, especially close to electronic degeneracies\cite{Schmerwitz2022, Ivanov2021, Levi2020b, Levi2020a} (a similar improvement has been established before for direct minimization algorithms in the context of ground state calculations\cite{Voorhis2002}). DO has for example been used to calculate the potential energy surface of excited states around an avoided crossing and conical intersection in the ethylene molecule, demonstrating that good results can be obtained even with semi-local generalized gradient approximation (GGA) functionals\cite{Schmerwitz2022}. More recently, the approach has been applied in density functional calculations of a large set of Rydberg excited states of molecules using GGA, meta-GGA as well as self-interaction corrected functionals\cite{Sigurdarson2023}. DO has also been used in calculations of excited states of solid-state systems with quantum point defects\cite{Ivanov2023, Sajid2023, Bertoldo2022}. 

In the present article, the DO method is used to perform orbital optimized calculations on a large set of charge transfer excited states in organic molecules. The main aim is to assess the accuracy of calculated excitation energy values for intramolecular charge transfer states in comparison to TD-DFT when using computationally inexpensive local and semi-local density functionals. Previous studies have focused on core\cite{Hait2020core} and Rydberg\cite{Sigurdarson2023, Seidu2015} excited states of molecules, or charge transfer states in bimolecular systems or large covalently linked dimeric complexes, where donor and acceptor are separated by a large distance\cite{Hait2021, Zhao2019, Barca2018, Briggs2015, Dreuw2004}. Local and semi-local GGA functionals developed for ground state KS calculations have been shown to give remarkably good results in such cases, with absolute errors on the vertical excitation energy relative to experimental and theoretical best estimates typically below 0.5\,eV. This performance represents a significant improvement over TD-DFT calculations, which can yield errors of several eV for high-energy and charge transfer excitations. However, it is still not clear how orbital optimized calculations with local and semi-local functionals perform compared to TD-DFT in the case charge transfer happens on shorter distances within the same chromophore, despite this type of excitation being ubiquitous in applications for solar energy conversion. 

Calculations with the local density approximation (LDA)\cite{Perdew1992} and with the GGA functionals PBE\cite{Perdew1996} and BLYP\cite{Becke1988, Lee1988} are carried out for 27 excitations in 15 organic molecules where reference values of the excitation energy are available from previous calculations using highly accurate wave function methods\cite{Loos2021}. A charge transfer distance measuring the spatial extent of charge separation is computed using the difference between the ground and excited state electron density obtained from both orbital optimized and TD-DFT calculations\cite{LeBahers2011}. TD-DFT is generally found to give larger charge transfer distances compared to the time-independent approach for all local and semi-local functionals considered here. In some cases, TD-DFT gives too low charge transfer distances because of mixing with Rydberg states when the basis set used includes diffuse functions, an issue that is shown here not to occur in the orbital optimized calculations. Both the TD-DFT and orbital optimized calculations are found to underestimate the excitation energy. However, while a strong correlation of the accuracy with the extent of charge transfer is observed for TD-DFT,  no correlation with the strength of charge transfer is found for the orbital optimized approach. The most significant improvements are obtained for the excitations with long charge transfer distance, where the absolute errors can be as large as 2\,eV for the TD-DFT calculations with local and semi-local functionals. However, the mean absolute error of the orbital optimized calculations over the full set of excitations remains relatively large, $\sim$0.7 eV. Preliminary calculations using the B3LYP\cite{Stephens1994, Becke1993, Lee1988} and CAM-B3LYP\cite{Yanai2004} functionals show that the charge transfer distance changes only slightly upon the inclusion of exact exchange and long-range correction, while the excitation energy values are improved for both short- and long-range charge transfer states.  

\section{Methodology}
\subsection{Direct orbital optimization}
In state-specific electronic structure methods, ground and excited states can be found as stationary points on the electronic energy landscape, the surface described by the variation of the energy as a function of the electronic degrees of freedom\cite{MolecularElectronicStructureTheory}. The ground state corresponds to a global minimum, while excited states are typically saddle points\cite{Schmerwitz2023, Marie2023, Burton2022, Kossoski2022, Perdew1985}. Within the KS method\cite{Kohn1965}, the excited stationary states correspond to higher-energy solutions of the KS equations and have non-aufbau orbital occupations. Therefore, they can be found by solving the KS equations through SCF algorithms based on sequential eigendecomposition of the KS Hamiltonian matrix if a given non-aufbau occupation of the orbitals can be preserved during the iterations. DO\cite{Schmerwitz2023, Ivanov2021, Levi2020b} is an alternative to conventional SCF approaches\cite{Lehtola2020} where a stationary solution is obtained by directly finding a unitary transformation of the orbitals that makes the energy stationary. Therefore, DO represents an extension of ground state direct minimization methods\cite{Voorhis2002, Hutter1994, Head-Gordon1988} to saddle points.

In the linear combination of atomic orbitals (LCAO) representation, a set of $M$ initial (occupied and unoccupied) molecular orbitals $\{\psi_i(\vect{r}) \vert i=1, 2, \dots, M\}$ is described as a linear combination of $M$ basis functions $\{\chi_\mu(\vect{r}) \vert \mu=1, 2, \dots, M\}$
\begin{equation}
    \centering
    \psi_i(\vect{r}) = \sum_{\mu=1}^M C_{\mu i} \chi_\mu(\vect{r})\,.
\end{equation}
The coefficients $O_{\mu i}$ of the orbitals that make the energy stationary, the optimal orbitals, can be found through sequential application of a unitary transformation of the reference coefficients $C_{\mu i}$
\begin{equation}
    \centering
    O_{\mu i} = \sum_{j=1}^M C_{\mu j}U_{ji} = \sum_{j=1}^M C_{\mu j}\left[ e^{\boldsymbol{\kappa}} \right]_{ji}\,,
\end{equation}
where the $M \times M$ unitary matrix $\matr{U}$ has been parameterized through the exponential of an anti-Hermitian matrix $\boldsymbol{\kappa} = - \boldsymbol{\kappa}^{\dagger}$. Thus, the energy is a function of the elements of $\boldsymbol{\kappa}$ representing orbital rotation angles. Since the anti-Hermitian matrices $\boldsymbol{\kappa}$ form a linear space, stationary points of the energy can be found using standard numerical optimization techniques, such as quasi-Newton methods. More details on the evaluation of the matrix exponential and the energy gradient can be found elsewhere\cite{Ivanov2021, Ivanov2021cpc, Levi2020b}.

The quasi-Newton step is typically preconditioned using the inverse of the following diagonal approximation to the electronic Hessian\cite{Ivanov2021cpc, Head-Gordon1988}
\begin{equation}\label{eq:precond}
    \centering
    \frac{\partial^{2}E}{\partial \kappa_{ij}^{2}} \approx -2 \left( \epsilon_i- \epsilon_j \right) \left( f_i - f_j\right)\,,
\end{equation}
where $\epsilon_i$ and $f_i$ are the energy and occupation numbers of the canonical orbitals of the initial guess for the calculation, respectively. 

\subsubsection{Freeze-and-release direct optimization}\label{sec:freeze_and_release}
Typically, the initial guess for a calculation of a single-electron excitation is constructed by performing a 90$^\circ$ rotation between an occupied orbital $\psi_a$ and an unoccupied orbital $\psi_r$ of the ground state corresponding to swapping the occupation numbers of the two orbitals. For example, for a LUMO$\leftarrow$HOMO excitation, a 90$^\circ$ rotation between the HOMO and the LUMO is performed, meaning that in the initial guess a hole is left in the ground state HOMO (occupation number of 0) and an electron is placed in the ground state LUMO (occupation number of 1). In the DO-MOM method, MOM is employed to restore the non-aufbau occupation of the orbitals in case of variational collapse during the electronic optimization. However, this approach has been shown not to be able to prevent variational collapse to unphysical, charge-delocalized solutions for some intramolecular charge transfer excitations\cite{Schmerwitz2023}.

Here, we adopt an alternative strategy where an initial constrained optimization is performed where the orbital of the hole, $\psi_a$, and the orbital of the excited electron, $\psi_r$, are frozen, meaning all pairwise orbital rotations involving these orbitals are constrained. This constrained optimization can be performed by applying the exponential transformation only in the subspace of $N=M-2$ orbitals. Let $l \in \left\{1, 2,\dots,N\right\}$ denote the orbital index in the subspace containing the $N$ orbitals with indices $s_{l}$. Then, the coefficients of the selected $N$ unconstrained orbitals are $C^\prime_{\mu l} = C_{\mu i}$, $i = s_{l}$\,, leading to the partly relaxed orbitals
\begin{align}
\tilde{C}_{\mu i} =
\begin{cases}
    \sum_{k=1}^N C^{\prime}_{\mu k}\left[ e^{\boldsymbol{\kappa}^\prime}\right]_{kl} \quad
    &\mr{for} \quad i = s_l \in \{s_1, \dots, s_N\}
    \\
    C_{\mu i} \quad
    &\mr{for} \quad i \notin \{s_1, \dots, s_N\}
\end{cases}\,,
\end{align}
where $\boldsymbol{\kappa}^\prime$ includes only rotations within the subspace of the $N$ unconstrained orbitals. This step of constrained optimization corresponds to a minimization because the degrees of freedom corresponding to negative curvature on the electronic energy surface are constrained.
The constraints are released, and a new preconditioner is computed according to eq \ref{eq:precond}. This updated preconditioner provides an improved estimation of the directions of negative curvature for the final step of unconstrained optimization in the space of all orbitals. In the following, this strategy is referred to as freeze-and-release DO. 

\subsection{Computational settings}
Time-independent, orbital optimized density functional and TD-DFT calculations are carried out for 27 charge transfer excited states in 15 organic molecules for which reference excitation energy values have previously been obtained using high-level wave function calculations\cite{Loos2021}. The calculations use molecular geometries obtained in ref \citen{Loos2021} by optimizing the ground state atomic structure in vacuum with the coupled cluster method at the CCSD(T) or CC3 level of theory. For two of the molecules, dimethylaminobenzonitrile (DMABN) and $N$-Phenylpyrrole (PP), excited states for two geometries, a planar and a twisted one, are calculated. 

Orbital optimized and the TD-DFT calculations with the LDA as well as the two commonly used GGA functionals PBE and BLYP are performed on the entire set of excitations. In addition, calculations with the global hybrid functional B3LYP, which includes 20\% of exact exchange, and the range-separated hybrid functional CAM-B3LYP, which adds 65\% exact exchange in the long range, are carried out for selected excitations, which are not affected by the presence of delocalized solution (see section \ref{sec:selecting_solutions}). 

All TD-DFT calculations are performed with version 5.0 of the ORCA software\cite{Neese2012, Neese2022} and use the linear-response and adiabatic approximations, with all the electrons (core and valence) represented with the aug-cc-pVDZ basis set. The orbital optimized calculations with the LDA and the GGA functionals are performed with the Grid-based Projector Augmented Wave (GPAW) software\cite{Mortensen2023, Enkovaara2010, Mortensen2005}, making use of the frozen core approximation and the projector augmented wave (PAW) formalism\cite{Blochl1994}. For these calculations, the valence electrons are represented in an LCAO basis consisting of primitive Gaussian functions taken from the aug-cc-pVDZ set\cite{Dunning1989, Kendall1992, Woon1994} augmented with a single set of numerical atomic orbitals (Gaussian basis set + sz)\cite{Rossi2015, Larsen2009}. The orbitals are represented on a uniform grid with a spacing between the points of 0.15\,\AA. The size of the simulation cell includes at least 8\,\AA\ of vacuum between any given atom and the closest edge of the cell. A test on a representative excitation shows that the effect of using the frozen core approximation and the PAW correction compared to all-electron calculations is negligible (see the Supporting Information). The freeze-and-release DO strategy illustrated in section \ref{sec:freeze_and_release} is employed, using a limited-memory SR1 algorithm\cite{Levi2020b} with a maximum step size of 0.2 for the constrained minimization step and 0.1 for the full optimization after releasing the constraints. Unless otherwise stated, the GPAW calculations are converged to a precision of $4 \cdot 10^{-8}\,\mathrm{eV}^{2}$ per valence electron in the squared residual of the KS equations. The orbital optimized calculations with the B3LYP and CAM-B3LYP hybrid functionals are performed with the ORCA software using the aug-cc-pVDZ basis set for all the electrons. For these calculations, a DIIS algorithm\cite{Kollmar1996} with level shifting of 1.5 Ha is used. For all orbital optimized calculations, no orthonormality constraints to lower-energy states are used, so the calculations are fully variational. The Supporting Information shows that the excitation energy is converged to within 10 meV with respect to the basis set.

All excited states considered here are singly excited open-shell singlets. For the orbital optimized calculations, the excited state energy is computed using the spin purification formula\cite{Ziegler1977}
\begin{equation}
    \centering
    E_{\mr s} = 2E_{\mr m} - E_{\mr t}\,,
    \label{eq:spin} 
\end{equation}
where $E_{\mr m}$ and $E_{\mr t}$ are the energy of the mixed-spin and triplet solutions obtained by variationally optimizing the orbitals in two separate calculations. The first one is initialized by promoting one electron from an occupied to an unoccupied ground state orbital within one spin channel, and the other is initialized by promoting the electron between spin channels. The orbitals used as initial guess for an orbital optimized excited state calculation are obtained from a ground state calculation at the same level of theory. The orbitals involved in the excitation, which are constrained in the initial step of the freeze-and-release DO strategy for the orbital optimized calculations with the LDA and the GGA functionals, are chosen as the orbitals of the electron-hole pair that contributes most significantly to the excited state obtained in the corresponding linear-response TD-DFT calculation. The target intramolecular charge transfer excited states are chosen as the ones obtained in the linear-response CCSD calculations reported in ref \citen{Loos2021}, which are taken here as a reference for assessing the accuracy of the calculated excitation energy. 

To characterize the extent of charge transfer in the orbital optimized and TD-DFT calculations, a charge transfer distance $d^{\mathrm{CT}}$ is calculated according to the metric introduced by Le Bahers et al.\cite{LeBahers2011} 
\begin{equation}
    \centering
    d^{\mathrm{CT}} = \dfrac{\left| \int \Delta \rho(\boldsymbol{r}) \boldsymbol{r} \, \mr{d}\boldsymbol{r} \right|}{q^{\mathrm{CT}}}\,
    \label{eq:dct} 
\end{equation}
where $\Delta \rho(\boldsymbol{r})=\rho_{\mr e}(\boldsymbol{r})-\rho_{\mr g}(\boldsymbol{r})$ is the difference between the real space excited and ground state electron densities, $q^{\mathrm{CT}}$ is the transferred charge defined as the integral over all space of the positive part of $\Delta \rho(\boldsymbol{r})$, and $\int \Delta \rho(\boldsymbol{r}) \boldsymbol{r} \, \mr{d}\boldsymbol{r}$ corresponds to the difference between the electric dipole moments of the ground and excited states. A charge transfer distance proportional to the difference between dipole moments is a better measure of the extent of charge transfer than the dipole moment of the excited state, since the movement of charge can be from a region with excess negative charge to a region with excess positive charge, thereby lowering the dipole moment of the molecule. Since $d^{\mathrm{CT}}$ represents the distance between the centroids of the positive and negative parts of the electron density difference, Le Bahers' metric defines the distance of a unidirectional displacement of charge. The excitations considered in the present work have a prevalent unidirectional character, in which case the charge transfer distance computed according to eq \ref{eq:dct} has been found to be in good agreement with more advanced metrics, which take the density shift in all directions into account\cite{wang2023earth}. For the time-independent calculations, the difference density, $\Delta \rho(\boldsymbol{r})$, is calculated as the difference between the fully relaxed electron density of the spin-mixed excited state solution and the ground state density. For linear-response TD-DFT, the one-particle difference density matrix from which $\Delta \rho(\boldsymbol{r})$ is obtained consists of two parts\cite{Pastore2017, Ipatov2009}. The first part, often called unrelaxed difference density matrix, is obtained from the eigenvectors of the Casida equations and includes only occupied-occupied and virtual-virtual orbital contributions. The second part contains occupied-unoccupied orbital contributions and is obtained by solving the so-called Z-vector equations\cite{Furche2002, Handy1984}. In the present work, the electron density obtained from the full TD-DFT one-particle difference density matrix is referred to as relaxed difference density, as is commonly done\cite{Loos2021, Pastore2017, Ipatov2009}.

\subsection{Characterizing and selecting multiple orbital optimized solutions}\label{sec:selecting_solutions}
\begin{figure}[hbt]
    \centering
    \includegraphics[width=0.75\textwidth]{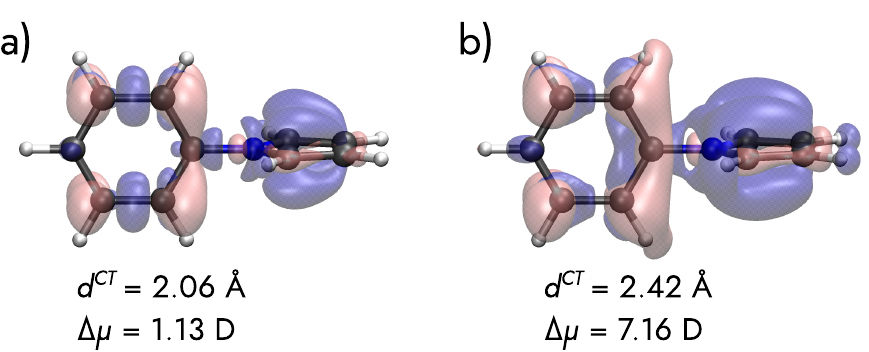}
    \caption{Difference between ground and excited state electron densities for the A$_1$ excitation of the twisted $N$-Phenylpyrrole molecule for (a) the more charge-delocalized solution and (b) the more charge-localized solution, as obtained from orbital optimized calculations with the PBE functional started by promoting an electron from an orbital localized on the pyrrole group to an orbital localized on the phenyl group. Red and blue isosurfaces represent the negative and positive parts of the density difference, respectively, and correspond to an isovalue of 0.015 Å$^{-3}$. The charge transfer distance, $d^{\mathrm{CT}}$, and the magnitude of the difference between the ground state and excited state dipole moments, $\Delta \mu$, are indicated for both solutions. TD-DFT calculations with with the range-separated CAM-B3LYP functional give a charge transfer distance and change in dipole moment of 2.41 Å and 7.95 D, respectively.}
   \label{fig:pp_tw_diff}
\end{figure}

Due to the non-linearity of the orbital optimization, orbital optimized calculations can be affected by the presence of stationary points on the electronic energy surface without a clear association with a particular electronic state\cite{Marie2023, Schmerwitz2022, MolecularElectronicStructureTheory}. In the context of intramolecular charge transfer excitations, it has been shown\cite{Schmerwitz2023} that orbital optimized calculations can converge on solutions with significantly different degree of charge transfer depending on the method used. This is exemplified by the A$_1$ LUMO+1$\leftarrow$HOMO charge transfer excitation in the twisted PP molecule\cite{Schmerwitz2023}. A calculation of the spin-mixed solution for the A$_1$ excited state of twisted PP  using the PBE functional is initialized by promoting an electron from the ground state HOMO localized on the pyrrole group to the ground state LUMO+1 localized on the phenyl group. A calculation with DO and MOM converges to a solution corresponding to a 2{\textsuperscript{nd}}-order saddle point with an excitation energy of 4.61\,eV. The freeze-and-release DO approach, instead, converges to a higher-energy solution corresponding to a 10\textsuperscript{th}-order saddle point with an excitation energy of 5.56\,eV (5.61 eV after spin purification using eq \ref{eq:spin}), much closer to the theoretical best estimate of 5.65\,eV \cite{Loos2021}. Figure \ref{fig:pp_tw_diff} shows the electron density difference between ground and excited states, for the two solutions of the A$_1$ excitation in the twisted PP molecule.
The solution obtained with freeze-and-release DO has larger dipole moment and charge transfer distance (9.36 D and 2.42 Å) compared to the solution obtained with DO-MOM (3.33 D and 2.06 Å). The dipole moment and charge transfer distance of the more charge-localized solution are in good agreement with the dipole moment and charge transfer distance obtained from the relaxed density of a TD-DFT calculation with the range-separated CAM-B3LYP functional (10.16 D and 2.41 Å), a functional that is known to provide an accurate description of intramolecular charge transfer excitations within TD-DFT\cite{Mester2022, Loos2021, Dev2012}. Clearly, the orbital optimized solution that best describes the target charge transfer excited state is the one obtained in the freeze-and-release DO calculations, while DO-MOM collapses to a solution where the transferred charge is too delocalized. This delocalization is caused by mixing of the hole originally localized on the pyrrole group with a lower-energy occupied orbital initially localized on the phenyl group\cite{Schmerwitz2023}.

Multiple solutions can further be characterized by a distance in number of electrons from the initial guess\cite{Barca2018jctclett, Thom2008}
\begin{equation}
    \centering
    \eta = N - \sum_{ab} \left| \Braket{ \psi_a^0 | \psi_b } \right|^2,
    \label{eq:eldist} 
\end{equation}
where $\psi_a^0$ and $\psi_b$ are the occupied orbitals of the initial guess and the converged solution, respectively. According to eq \ref{eq:eldist}, $\eta$ can have values between 0 and $N$. The closer to zero the distance is, the more closely the solution resembles the initial guess. Therefore, $\eta$ is a simple measure of the extent of orbital relaxation in the excited state. Too large values may signify a variational collapse. In the case of the PBE orbital optimized calculations of the twisted PP molecule, the more charge-delocalized solution has a relatively large electronic distance from the initial guess of 0.50, compared to the charge-localized solution ($\eta$=0.27). 

When multiple solutions for a given excitation are found, the one providing an excitation energy and charge transfer distance closer to the values obtained in reference calculations is selected for the assessment of the accuracy of the orbital optimized calculations as presented below. Other solutions with an excitation character different from the target solution and without a clear association to a particular electronic state are deemed unphysical. 

\section{Results}
Table \ref{tbl:energies} reports the excitation energy values of all intramolecular charge transfer states obtained here using the LDA, and the PBE and BLYP functionals in orbital optimized and TD-DFT calculations. The charge transfer distance, $d^{\mathrm{CT}}$, computed from the electron density of the orbital optimized spin-mixed excited state solution obtained with the PBE functional as well as the distance, $\eta$, of this solution from the initial guess is also reported.
\begin{table}[hbtp]
\scriptsize
  \caption{Excitation energy (in eV) of charge transfer states of organic molecules obtained from orbital optimized calculations using DO and TD-DFT calculations with local and semi-local functionals, together with the mean absolute and mean signed errors (MAE and MSE) with respect to theoretical best estimate (TBE) values\cite{Loos2021}. The charge transfer distance (in Å), $d^{\mathrm{CT}}$, of the orbital optimized spin-mixed excited state solution obtained with PBE is also shown together with the electronic distance, $\eta$, of this solution from the initial guess.}
  \label{tbl:energies}
  \begin{tabular}{lc|r|rrr|rrrrr}
    \hline
    Molecule  & Sym. & \multicolumn{1}{c}{TBE$\,^\textrm{a}$} & \multicolumn{3}{|c}{TD-DFT} & \multicolumn{5}{|c}{Orbital optimized} \\
     \hline
   & & & \multicolumn{1}{c}{LDA} & \multicolumn{1}{c}{BLYP} & \multicolumn{1}{c|}{PBE} &  \multicolumn{1}{c}{LDA} & \multicolumn{1}{c}{BLYP} & \multicolumn{3}{c}{PBE}  \\
   \cline{9-11} 
   & & &  &  &  &   &  & \multicolumn{1}{c}{$\Delta$E} & \multicolumn{1}{c}{$d^{\mr{CT}}$}  & \multicolumn{1}{c}{$\eta$} \\
    \hline
    Aminobenzonitrile (ABN)  & A$_1$ & 5.09 & 4.65	& 4.61 & 4.65 &	3.97 & 3.98 &	4.01  & 1.06 & 0.03  \\
    Aniline   & A$_1$ & 5.48 & 5.15 & 5.07 & 5.13 & 4.66 & 4.67 & 4.72  & 0.82 & 0.02 \\
    Azulene   &  A$_1$ & 3.84 & 3.48 & 3.43 & 3.47 & 3.32 & 3.35 & 3.39  & 0.94 & 0.03 \\
       & B$_2$  & 4.49 & 4.45 & 4.41 & 4.45 & 4.17 & 4.09 & 4.12   & 0.77 & 0.03 \\
    Benzonitrile   & A$_2$  & 7.05 & 5.75 & 5.75 & 5.76 & 6.87 & 6.06 & 6.66$\,^\textrm{c}$ & 1.04 & 0.49 \\
    Benzothiadiazole (BTD) & B$_2$  & 4.28 & 3.58 & 3.52 & 3.57 & 3.41 & 3.42 & 3.46  & 1.19 & 0.04 \\
    Dimethylaminobenzonitrile (DMABN) & A$_1$  & 4.86 & 4.31 & 4.32 & 4.34 & 3.77 & 3.79 & 3.80  & 1.56 & 0.05 \\ 
    Twisted DMABN &  A$_2$ & 4.12 & 2.17 & 2.36 & 2.30 & 3.59 & 3.62 & 3.55  & 2.04 & 0.12 \\
    & B$_1$  & 4.75 & 2.76 & 2.95 & 2.90 & 4.25 & 4.30 & 4.23  & 1.75 & 0.12 \\
    Dimethylaniline (DMAn) & B$_2$  & 4.40 & 3.94 & 4.01 & 4.01 & 3.98 & 3.97 & 3.99  & 1.08 & 0.06 \\
    &  A$_1$ & 5.40 & 4.89 & 4.93 & 4.97 & 4.46 & 4.49 & 4.52  & 1.33 & 0.04 \\
    Hydrogen Chloride & $\Pi$  & 7.88 & 7.16 & 6.89 & 7.10 & 7.78 & 7.48 & 7.55  & 0.86 & 0.07\\
    Nitroaniline & A$_1$  & 4.39 & 3.50 & 3.48 & 3.53 & 3.46 & 3.45 & 3.49  & 2.05 & 0.16 \\
    Nitrobenzene & A$_1$  & 5.39 & 4.36 & 4.31 & 4.38 & 4.45 & 4.45 & 4.53  & 1.46 & 0.11\\
    Nitrodimethylaniline (NDMA) & A$_1$  & 4.13 & 3.19 & 3.20 & 3.23 & 3.21 & 3.21 & 3.24  & 2.34 & 0.18 \\
    Nitropyridine $N$-Oxide (NPNO) & A$_1$  & 4.10 & 3.47 & 3.43 & 3.48 & 3.05 & 3.03 & 3.08  & 1.72 & 0.13\\
    $N$-Phenylpyrrole (PP) & B$_2$  & 5.32 & 4.01 & 4.01 & 4.03 & 4.41 & 4.45 & 4.51  & 1.59 & 0.20 \\
      & A$_1$ & 5.86  & 3.93 & 3.94 & 3.96 & 5.34 & 5.29 & 5.32  & 2.02 & 0.35 \\
    Twisted PP & B$_2$ & 5.58 & 3.56 & 3.58 & 3.59 & 5.41 & 5.29 & 5.29  & 2.36 & 0.15\\
    &  A$_1$ & 5.65 & 3.60 & 3.62 & 3.63 & 5.71 & 5.59 & 5.61  & 2.41 & 0.27 \\
    & A$_2$  & 5.95 & 4.16 & 4.17 & 4.19 & 5.50 & 5.43 & 5.41  & 2.15 & 0.21\\
    & B$_1$  & 6.17 & 4.19 & 4.21 & 4.22 & 5.57 & 5.59 & 5.59  & 2.16 & 0.27\\
    Phthalazine &  A$_2$ & 3.91 & 2.50 & 2.71 & 2.66 & 3.17 & 3.22 & 3.17  & 1.26 & 0.11\\
    &  B$_1$ & 4.31 & 3.10 & 3.29 & 3.25 & 3.66 & 3.68 & 3.64  & 1.26 & 0.07 \\
    Quinoxaline & B$_2$  & 4.63 & 3.73 & 3.71 & 3.75 & 3.78 & 3.81 & 3.85  & 1.25 & 0.05 \\
    & A$_1$  & 5.65 & 5.43 & 5.42 & 5.47 & 4.75 & 4.66 & 4.69  & 0.62 & 0.04\\
    & B$_1$  & 6.22 & 4.67 & 4.82 & 4.78 & 5.20 & 5.25 & 5.19  & 1.24 & 0.13 \\

    \hline
    \multicolumn{2}{l|}{MAE, all} & & 1.08 & 1.06 & 1.04 &  0.67 & 0.71 & 0.68  \\
    \multicolumn{2}{l|}{MAE, long-range only ($d^{\mathrm{CT}}_{\mathrm{OO}} > 1.5\,\text{\AA}$)} & & 1.50 & 1.47 & 1.46 & 0.64 & 0.65 & 0.65  \\
    \multicolumn{2}{l|}{MSE, all} & & -1.08 & -1.06 & -1.04 & -0.67 & -0.71 & -0.68  \\
    \multicolumn{2}{l|}{MSE, long-range only ($d^{\mathrm{CT}}_{\mathrm{OO}} > 1.5\,\text{\AA}$)} & & -1.50 & -1.47 & -1.46 & -0.64 & -0.66 & -0.66  \\ 
    \hline
    \vspace{-0.05cm}
  \end{tabular} \\
  \raggedright
 $^\textrm{a}$ Theoretical best estimates obtained at the CCSDT/aug-cc-pVQZ level in ref \citen{Loos2021} \\
 $^\textrm{c}$ Obtained for a squared residual of the KS equations of $2.5 \cdot 10^{-5}\,\mathrm{eV}^{2}$ per valence electron
\end{table}

The electronic distance from the initial guess is typically between 0.02-0.1 for states with $d^{\mathrm{CT}} < 1.5\,\text{\AA}$ and between 0.1-0.3 for states with $d^{\mathrm{CT}} > 1.5\,\text{\AA}$. The distance from the initial guess systematically increases with the charge transfer distance because the orbital relaxation effects are more pronounced. However, there are two cases where the distance from the initial guess is larger than expected based on this trend, the A$_2$ state of benzonitrile and the A$_1$ state of planar PP. For LDA and PBE orbital optimized calculations with an initial guess targeting the A$_2$ excited state of benzonitrile, two spin-mixed solutions are found. In the case of PBE, the first solution, obtained with DO-MOM, has $\eta=0.19$ and an excitation energy after spin purification of 6.02 eV. The charge transfer distance and dipole moment of this solution are 0.58 Å and 2.81 D, respectively. The second solution, obtained with freeze-and-release DO, is farther from the initial guess ($\eta = 0.49$), has an excitation energy of 6.66 eV, and charge transfer distance and dipole moment of 1.04 Å and 0.58 D, respectively (note that the charge transfer connected to this excitation lowers the dipole moment of the molecule). The theoretical best estimate of the excitation energy is 7.05 eV\cite{Loos2021}, while previous TD-DFT calculations with the range-separated hybrid functional CAM-B3LYP and CASSCF calculations\cite{Sobolewski1996} report a charge transfer distance and dipole moment of 1.17 Å and 0.64 D, respectively. Therefore, in this case it appears that the solution with largest distance from the initial guess is the one best describing the target charge transfer excitation. In the case of the A$_1$ excited state of planar PP ($\eta=0.35$ when using the PBE functional), the spin-mixed solution obtained with any of the three functionals exhibits greater delocalization of the transferred charge than the solution obtained for the A$_1$ state at the twisted geometry (see section \ref{sec:selecting_solutions}), despite previous calculations indicating a small dependence of the strength of charge transfer on the geometry\cite{Loos2021}. Any attempts using freeze-and-release DO and a DO approach based on generalized mode following (DO-GMF)\cite{Schmerwitz2023} could not find a spin-mixed excited state solution with enhanced charge localization closer to the initial guess for this state.  

The excitation energy values and charge transfer distances obtained for a subset of five charge transfer excited states from calculations using the global hybrid functional B3LYP and the range-separated hybrid CAM-B3LYP are reported in Table \ref{tbl:hybrids}.
\begin{table}[hbtp]
\small
  \caption{Excitation energy (in eV) and charge transfer distance $d^{\mathrm{CT}}$ (in Å) obtained from orbital optimized and TD-DFT calculations using the hybrid B3LYP and CAM-B3LYP functionals for a subset of five excited states, ranging from short- to long-range charge transfer states.}
  \label{tbl:hybrids}
  \begin{tabular}{lc|r|rrrr|rrrr}
    \hline
    Molecule  & Sym. & \multicolumn{1}{c}{TBE$\,^\textrm{a}$} & \multicolumn{4}{|c}{TD-DFT} & \multicolumn{4}{|c}{Orbital optimized} \\
     \hline
   & & & \multicolumn{2}{c}{B3LYP} & \multicolumn{2}{c|}{CAM-B3LYP} &  \multicolumn{2}{c}{B3LYP} & \multicolumn{2}{c}{CAM-B3LYP}   \\
   & & & $\Delta$E & $d^{\mr{CT}}$ & $\Delta$E & $d^{\mr{CT}}$ & $\Delta$E & $d^{\mr{CT}}$ & $\Delta$E & $d^{\mr{CT}}$ \\
    \hline
    Azulene & B$_2$ & 4.49 & 4.63 & 0.92 & 4.77 & 1.02 & 4.38 & 0.80 & 4.71 & 0.85 \\
    Quinoxaline &  B$_1$  & 6.22 & 5.69 & 1.21 & 6.47 & 1.24 & 5.82 & 1.22 & 6.34 & 1.22 \\
    Twisted DMABN &  B$_1$  & 4.75 & 3.84 & 1.79 & 4.69 & 1.74 & 4.49 & 1.70 & 4.60 & 1.70   \\
    NDMA & A$_1$  & 4.13 & 3.66 & 2.32 & 4.13 & 2.26 & 3.57 & 2.28 & 4.01 & 2.28 \\
    Twisted PP & A$_2$  & 5.95 & 4.96 & 2.49 & 5.89 & 2.35 & 5.66 & 2.29 & 5.80 & 2.37  \\
    \hline
    \end{tabular} \\
 $^\textrm{a}$ Theoretical best estimates obtained at the CCSDT/aug-cc-pVQZ level in ref \citen{Loos2021} \\
\end{table}
The states are chosen to be representative of the different extents of charge transfer in the full set of 27 excitations, ranging from short range to long range. The calculations performed for this subset of excited states are not affected by the presence of charge-delocalized solutions, for both local and hybrid functionals.

\subsubsection{Charge transfer distance}
A direct comparison of the calculated charge transfer distances to values obtained from higher-level calculations in previous works is not possible due to the use of different charge transfer metrics and/or the use of different basis sets, e.g., basis sets lacking diffuse functions, which can have a non-negligible effect on the charge transfer distances\cite{Loos2021}. In the following, a comparison is made between the charge transfer distances calculated with TD-DFT and those obtained from the orbital optimized calculations. Since the orbital optimized approach is fully variational, this comparison allows the assessment of relaxation effects in linear-response TD-DFT. The dependence of the charge transfer distances obtained with both TD-DFT and orbital optimized calculations with respect to the functional is also analyzed.

\begin{figure}[hbt]
    \centering
    \includegraphics[width=0.85\textwidth]{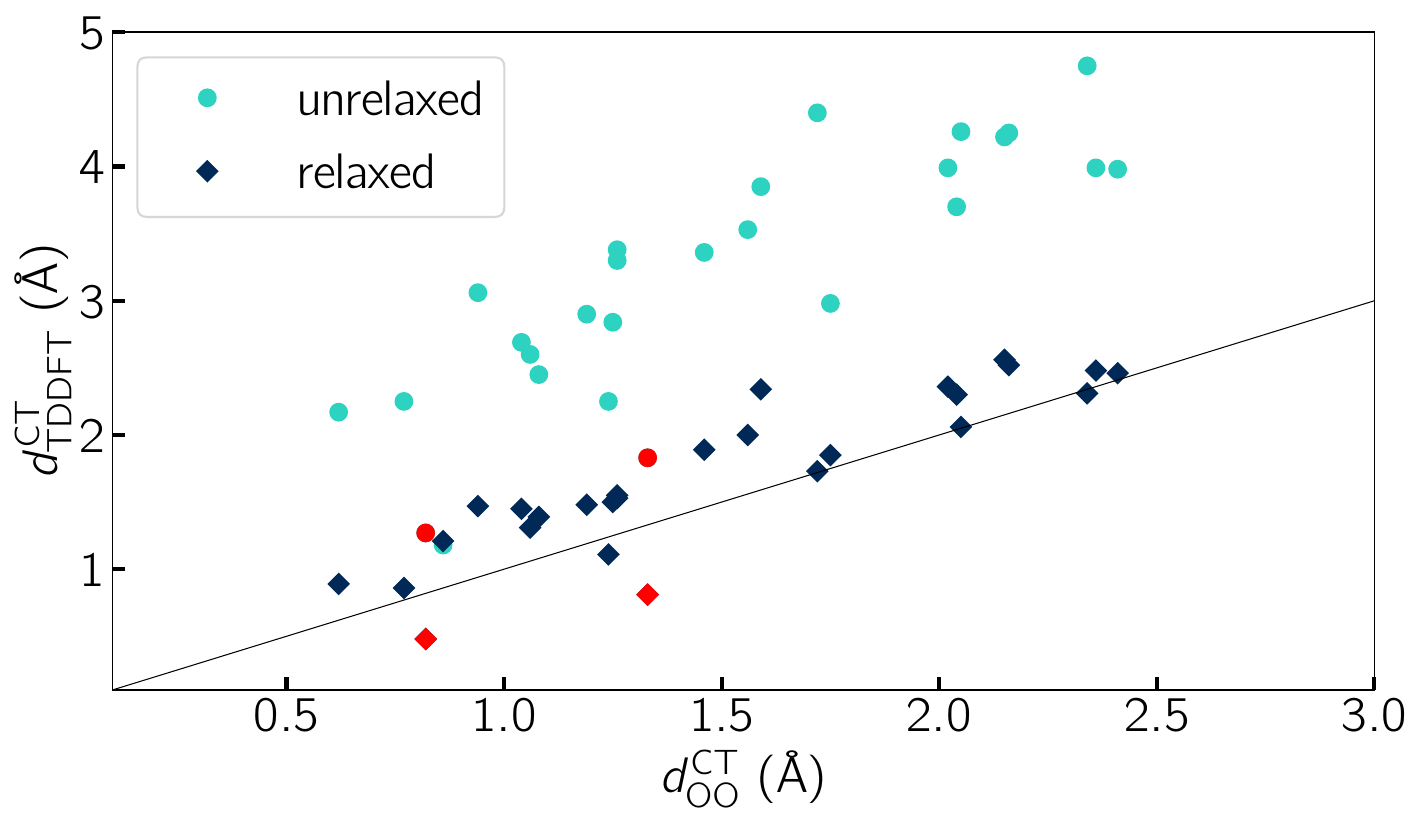}
    \caption{Comparison between charge transfer distances of excited states of organic molecules obtained from TD-DFT and orbital optimized calculations using the PBE functional. The black line represents one-to-one correspondence. TD-DFT overestimates the charge transfer distance both when using only the unrelaxed part of the difference density matrix and when employing the full, relaxed difference density matrix obtained through the Z-vector approach. The red points correspond to excitations where TD-DFT gives mixing with Rydberg states (see Figure \ref{fig:diffRyd}).}
    \label{fig:dct}
\end{figure}

Figure \ref{fig:dct} compares the charge transfer distances obtained with the PBE functional in the TD-DFT and orbital optimized calculations. 
When using only the unrelaxed part of the TD-DFT difference density matrix, the resulting charge transfer distances are significantly larger than those obtained from the orbital optimized calculations for all excitations. This is in agreement with the results of ref \citen{wang2023earth} for a core excitation of the water molecule, where relaxation effects are important. For charge transfer excitations, the electron density is expected to rearrange, reducing the electron-electron repulsion. This effect is clearly not captured by the unrelaxed TD-DFT density. Using the full, relaxed TD-DFT difference density matrix leads to better agreement with the orbital optimized calculations, but the charge transfer distance is still larger for most cases, indicating that relaxation effects are missing in linear-response TD-DFT even when employing the Z-vector method. These trends are the same for LDA and the BLYP functional (see the Supporting Information). 

\begin{figure}[hbt]
    \centering
    \includegraphics[width=0.5\textwidth]{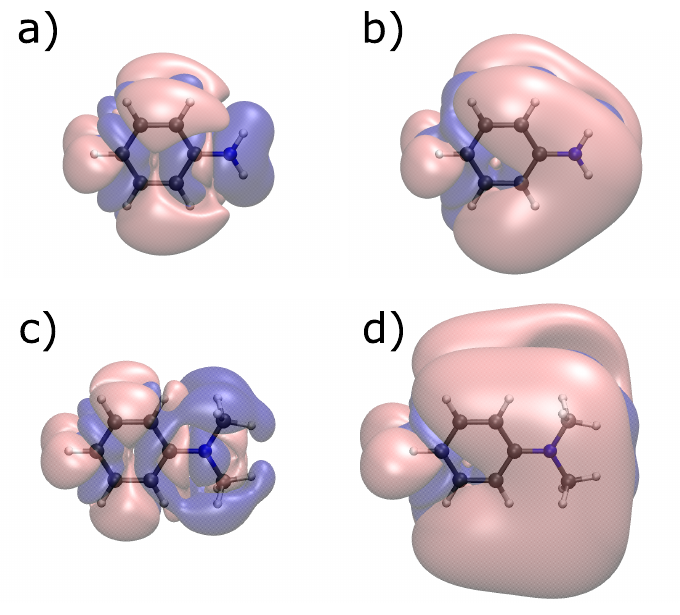}
    \caption{Difference between ground and excited state electron densities for the A$_1$ excitation of aniline (a, b) and dimethylaniline (c, d) as obtained from orbital optimized (a, c), and TD-DFT (b, d) calculations using the PBE functional. For the orbital optimized calculations, the difference density for a spin-mixed solution is visualized. Red and blue isosurfaces represent the negative and positive parts of the density difference, respectively, with isovalues including $\sim$90\% of the positive density difference. Contributions of Rydberg orbitals are observed in the TD-DFT calculations, whereas they are absent in the results of the orbital optimized calculations, even if the basis set used includes diffuse functions.}
   \label{fig:diffRyd}
\end{figure}

For a few excitations, the charge transfer distance obtained from TD-DFT is smaller than the charge transfer distance from the orbital optimized calculations. In most of these cases, the excitation is found to have partial Rydberg character in TD-DFT, while it has no Rydberg character in the orbital optimized calculations, as shown in Figure \ref{fig:diffRyd}.
The use of diffuse basis functions is known to lead to artificial mixing with Rydberg excitations in linear-response TD-DFT calculations\cite{Shu2017}. Indeed, if a cc-pVDZ+sz basis set is used, which does not include diffuse functions, no mixing with Rydberg states is observed in the TD-DFT calculations. This issue does not affect the orbital optimized calculations (see Figure \ref{fig:diffRyd}). 

\begin{figure}[hbt]
    \centering
    \includegraphics[width=0.85\textwidth]{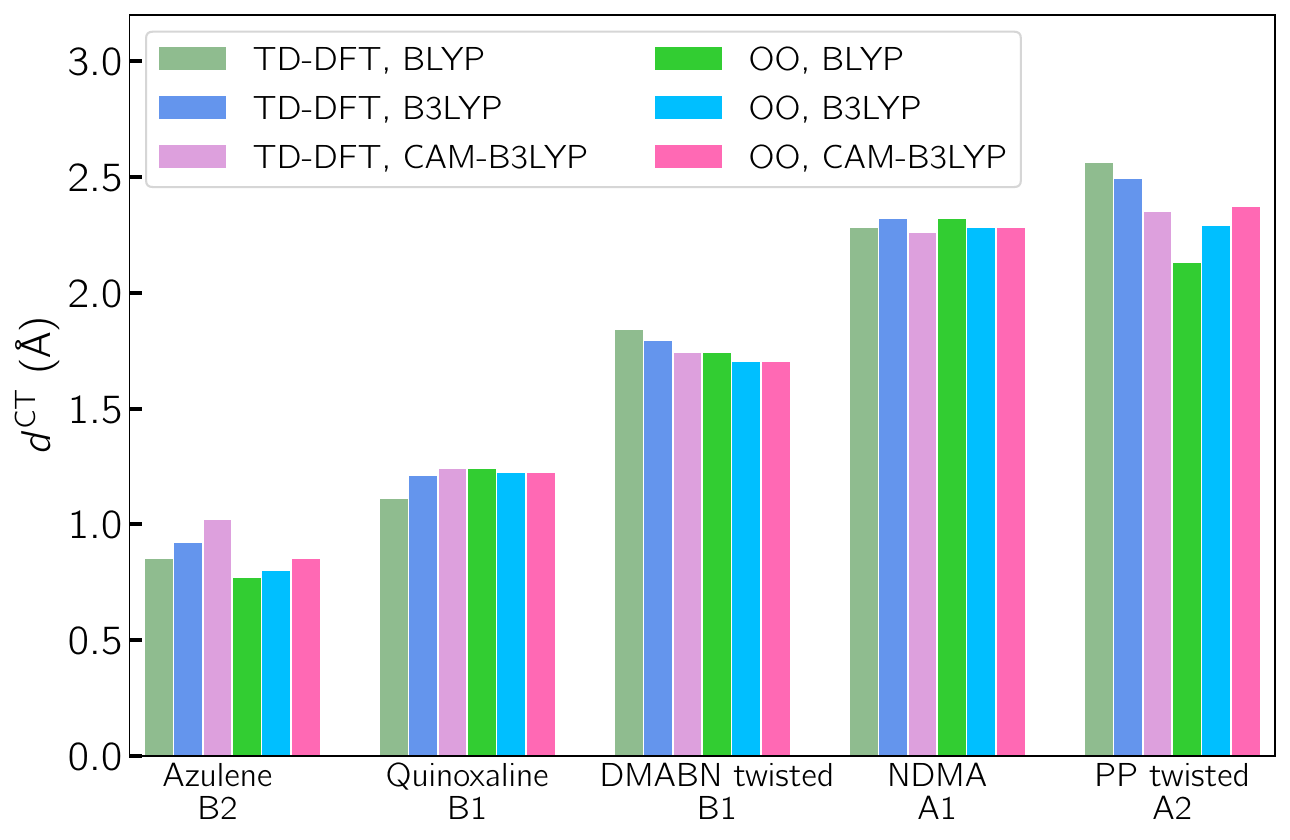}
    \caption{Charge transfer distances of selected excited states of organic molecules obtained from orbital optimized and TD-DFT calculations using the BLYP, B3LYP and CAM-B3LYP functionals. The TD-DFT values are calculated using the full, relaxed difference density matrix obtained through the Z-vector approach.}
   \label{fig:dct_h}
\end{figure}

A comparison between the charge transfer distances calculated from the TD-DFT relaxed difference density matrix and the orbital optimized calculations using the BLYP, B3LYP, and CAM-B3LYP functionals is shown in Figure \ref{fig:dct_h}. 
In most cases, the functional has a small effect on the charge transfer distance, especially for the orbital optimized calculations. Apart from the state with shortest charge transfer distance (B$_2$ state of azulene), the orbital optimized calculations using the range-separated hybrid functional CAM-B3LYP give charge transfer distances very close to those obtained from the TD-DFT calculations using CAM-B3LYP. 

\subsubsection{Excitation energy}
\begin{figure}[hbt]
    \centering
    \includegraphics[width=0.85\textwidth]{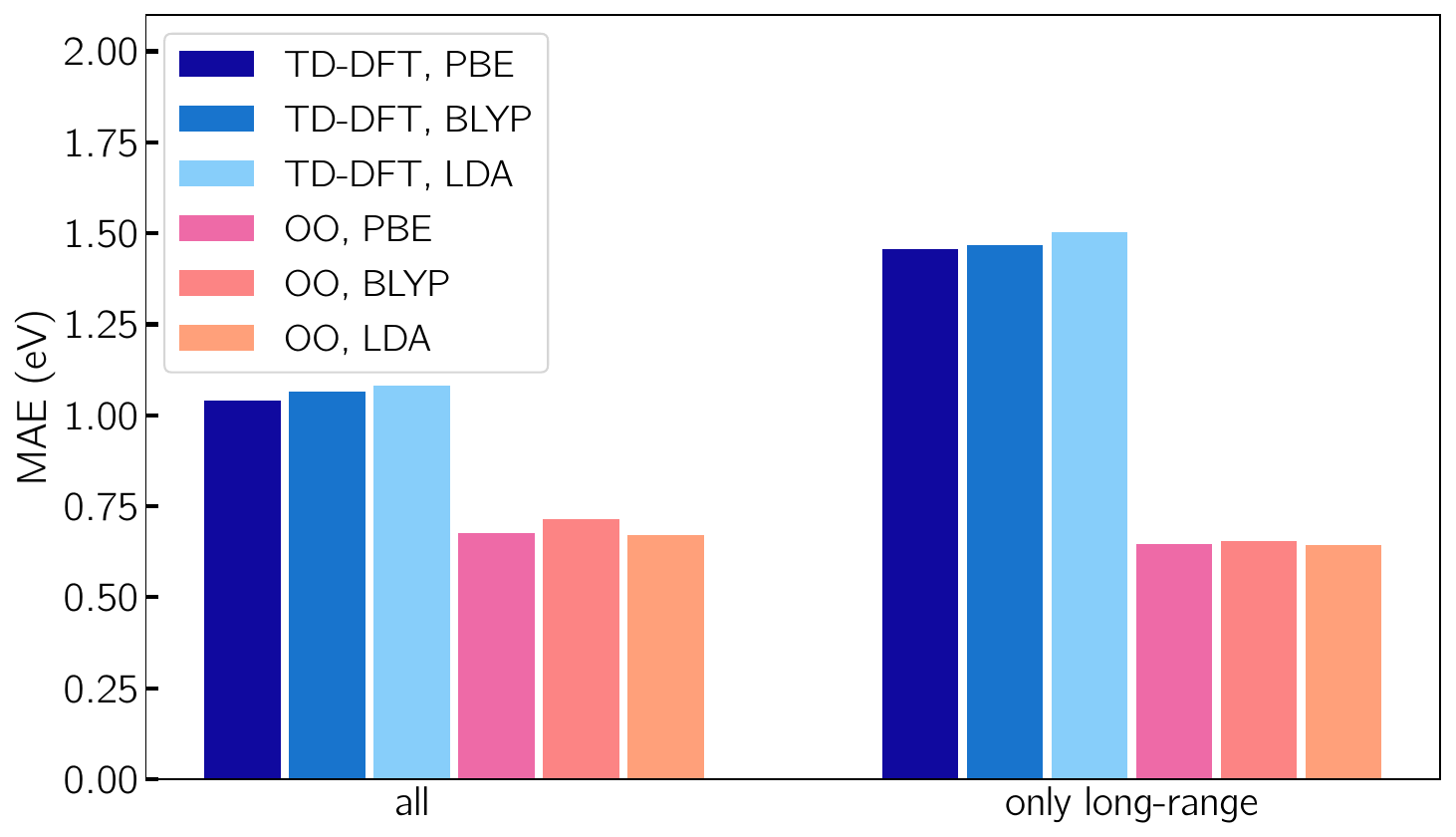}
    \caption{Mean absolute error (MAE) on the excitation energy of charge transfer states of organic molecules relative to theoretical best estimates\cite{Loos2021} for orbital optimized and TD-DFT calculations. The orbital optimized approach yields on average smaller errors than TD-DFT. The improvement is most significant for the long-range charge transfer excitations ($d^{\mathrm{CT}} > 1.5\,\text{\AA}$, as evaluated from the spin-mixed orbital optimized excited state solution).}
   \label{fig:mae}
\end{figure}

As Table~\ref{tbl:energies} shows, both the orbital optimized and TD-DFT calculations using local and semi-local functionals systematically underestimate the excitation energy compared to theoretical best estimates obtained at the CCSDT/aug-cc-pVQZ level\cite{Loos2021}. However, the errors of the orbital optimized calculations are on average lower than those of TD-DFT. Figure ~\ref{fig:mae} compares the mean absolute errors (MAEs) of the orbital optimized and TD-DFT calculations for all three local and semi-local functionals as evaluated considering all 27 excitations as well as a subset of long-range charge transfer excitations selected according to $d^{\mathrm{CT}}_{\mathrm{OO}} > 1.5\,\text{\AA}$. 
When considering all excitations, the MAE of the orbital optimized calculations is $\sim$0.7\,eV for all three functionals, approximately 0.3\,eV smaller than that of the TD-DFT calculations. When considering only the excitations with longest charge transfer distance, the MAE of the TD-DFT calculations increases to around 1.4\,eV, whereas the MAE of the orbital optimized calculations remains largely unchanged. Moreover a trend can be seen for the TD-DFT calculations, where the LDA yields larger errors than the GGA functionals, while for the orbital optimized calculations all the functionals produce the same MAE within 0.01\,eV.

\begin{figure}[hbt]
    \centering
    \includegraphics[width=\textwidth]{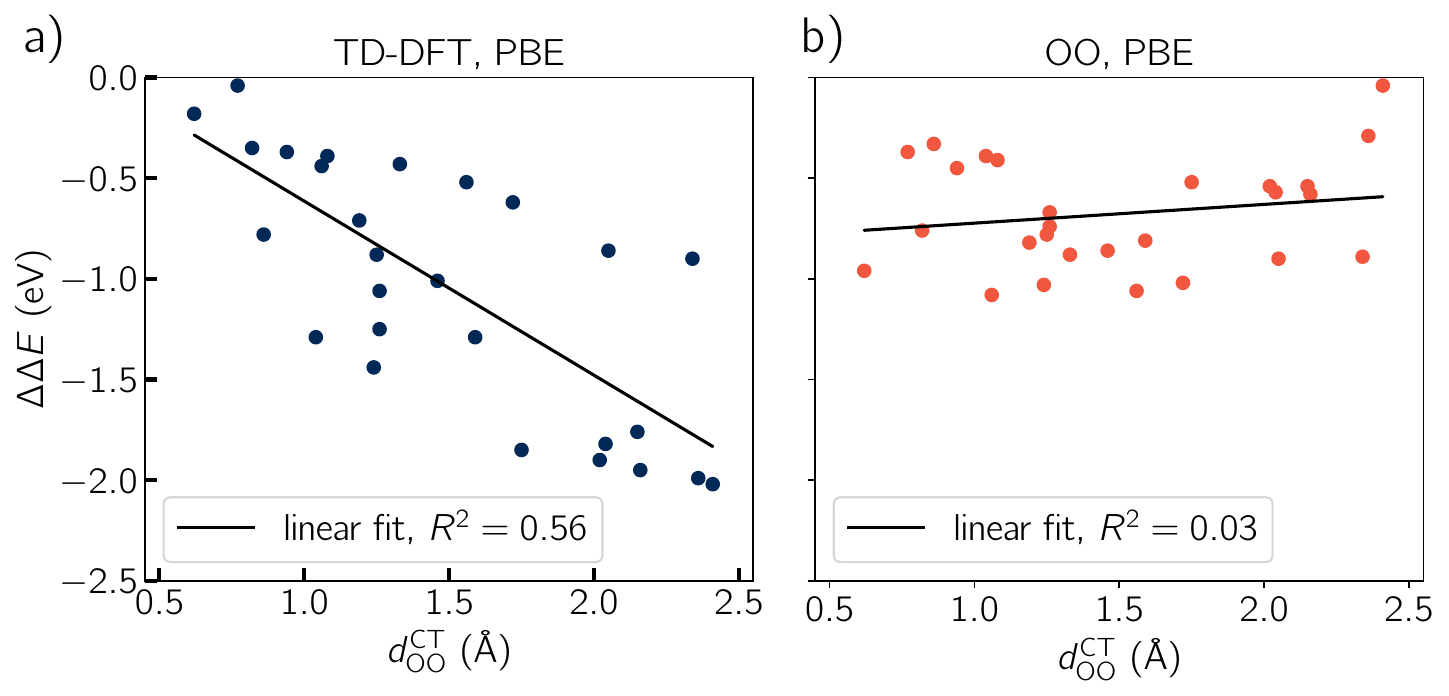}
    \caption{Error on the excitation energy of intramolecular charge transfer states of organic molecules relative to theoretical best estimates\cite{Loos2021} for TD-DFT and orbital optimized calculations as a function of the charge transfer distance, $d^{\mathrm{CT}}_{\mathrm{OO}}$. The calculations use the PBE functional and $d^{\mathrm{CT}}_{\mathrm{OO}}$ is evaluated from the spin-mixed orbital optimized excited state solution. The black lines represent linear regression fits. The error of TD-DFT increases with the charge transfer distance ($R^2$ of the fit of 0.56), while no correlation between the extent of charge transfer and the accuracy of the computed excitation energy is found for the orbital optimized calculations ($R^2$ of 0.03).}
   \label{fig:enscat}
\end{figure}

The dependence of the accuracy of the TD-DFT and orbital optimized calculations on the extent of charge transfer is further analyzed in Figure \ref{fig:enscat}, which plots the error on the excitation energy for the PBE functional as a function of the charge transfer distance obtained from the orbital optimized calculations, $d^{\mathrm{CT}}_{\mathrm{OO}}$.
As the extent of charge transfer increases, TD-DFT more significantly underestimates the excitation energy, yielding absolute errors up to $2$ eV for some of the excitations with longest charge transfer distance. As a consequence, TD-DFT predicts the A$_1$ states of planar PP with strong charge transfer character to be lower in energy than the more local B$_2$ state, which is in disagrement with high-level quantum chemistry calculations (see Table \ref{tbl:energies}). In contrast, the errors of the orbital optimized calculations are always within $-1.1$\,eV, and no negative slope with increasing charge transfer distance is observed. As a result, the order of the A$_1$ and B$_2$ states of planar PP is predicted correctly. These trends are the same for the LDA and the BLYP functional, as shown in the Supporting Information. A similar correlation between the underestimation of the excitation energy and the extent of charge transfer, though less pronounced, has been observed previously for TD-DFT calculations with the global hybrid functionals B3LYP and PBE0\cite{wang2023earth, Loos2021}, which include 20\% and 25\% of exact exchange, respectively. For the excitations with charge transfer distance bigger than 2.0 Å, the orbital optimized calculations with the LDA and GGA functionals (MAE $\sim$0.5\,eV) achieve better accuracy than TD-DFT with the more computationally expensive global hybrid functionals (MAE of 0.73\,eV and 0.91\,eV for B3LYP and PBE0, respectively\cite{Loos2021}). 

\begin{figure}[hbt]
    \centering
    \includegraphics[width=0.8\textwidth]{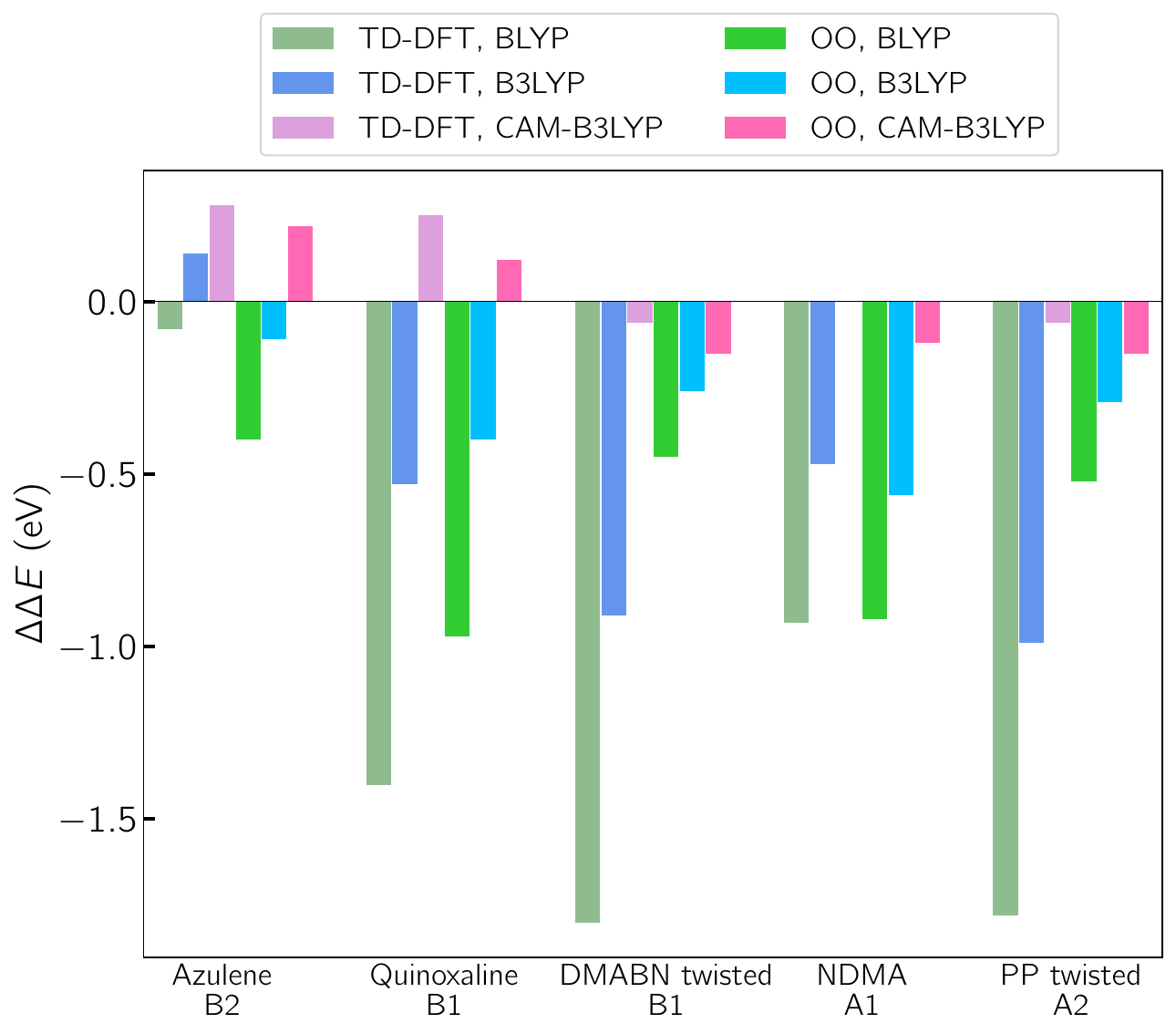}
    \caption{Error on the excitation energy of a subset of five charge transfer states relative to theoretical best estimates\cite{Loos2021} for orbital optimized and TD-DFT calculations using the BLYP, B3LYP, and CAM-B3LYP functionals.}
   \label{fig:en_h}
\end{figure}

The effect of including exact exchange in the functional is analyzed in Figure \ref{fig:en_h}, which shows the error on the excitation energy of a subset of five excited states for orbital optimized and TD-DFT calculations with the BLYP, B3LYP, and CAM-B3LYP functionals. 
Using the global hybrid functional B3LYP reduces the errors of both the orbital optimized and TD-DFT calculations. However, for the excitations with long-range charge transfer, the absolute errors of TD-DFT remain large, up to $1$ eV, while the absolute errors of the orbital optimized calculations are less than 0.6 eV. For the B$_1$ state of twisted DMABN, the errors of the TD-DFT calculations with the BLYP and B3LYP functionals are bigger than for the A$_1$ state of NDMAN, despite the latter having a longer charge transfer distance. This is because the spatial overlap between the orbitals involved in the B$_1$ excitation of twisted DMABN is significantly smaller than for the A$_1$ state of NDMAN. As expected\cite{Mester2022, Loos2021, Dev2012}, the TD-DFT calculations with the range-separated hybrid functional CAM-B3LYP provide small errors for the excited states with long-range charge transfer but overestimate the excitation energy of the excited state with short charge transfer distance. An improvement of the excitation energy of long-range charge transfer states is also observed for the orbital optimized calculations with CAM-B3LYP compared to BLYP and B3LYP. There, the errors are bigger than for TD-DFT, but still relatively small ($\sim$-0.15 eV). Moreover, the orbital optimized calculations with CAM-B3LYP overestimate the excitation energy of the two short-range charge transfer states slightly less than TD-DFT. 

\section{Discussion and conclusions}
State specific orbital optimization in time-independent density functional calculations of excited states offers a distinct advantage over TD-DFT, particularly for describing high-energy and charge transfer excitations. However, orbital optimized calculations face two main challenges: (1) The excited state solutions are typically saddle points on the electronic energy surface, and (2) the non-linearity of the orbital optimization leads to a large number of stationary solutions, some lacking a clear association with a specific electronic state. 
Here, it is found that the presence of charge-delocalized solutions affects calculations of intramolecular charge transfer states with both local, semi-local, and hybrid functionals. Exploratory orbital optimized calculations with the B3LYP functional and using a DIIS-based approach have been performed for the A$_1$ excited state of the twisted PP molecule where a charge-localized and a charge-delocalized solution are found for the local and semi-local functionals (see section \ref{sec:selecting_solutions}). Similarly to the DO-MOM calculations with the PBE functional illustrated in section \ref{sec:selecting_solutions}, the B3LYP calculations converge to a mixed-spin solution where mixing between the hole initially localized on the pyrrole group and an occupied orbital initially localized on the phenyl group of the molecule leads to charge delocalization. This solution has a dipole moment and charge transfer distance of 3.82 D and 2.06 Å, respectively, much smaller than those of the charge-localized solutions obtained for the calculations with local and semi-local functionals (7.3-9.5 D and 2.2-2.4 Å, respectively), the latter being in agreement with the TD-DFT calculations using the CAM-B3LYP functionals (dipole moment and charge transfer distance of 9.56 D and 2.35 Å, respectively). 

The issue of multiple orbital optimized solutions has been observed for other classes of excitations. For example, it has recently been shown that the presence of charge-delocalized solutions affects orbital optimized density functional calculations of core excitations in symmetric molecules\cite{Hait2020core}, such as CO$_2$ and N$_2$, and charge transfer states of systems, such as H$_2$O-Li$^+$, where the donor and acceptor are separated by a large distance, even when global or range-separated hybrid functionals are used\cite{Hait2021}. Convergence on such unphysical solutions can have detrimental effects on the calculation of excited state potential energy surfaces and molecular dynamics simulations, therefore care must be taken to avoid converging on spurious solutions where the charge is too delocalized. Here, a freeze-and-release DO approach involving a first constrained optimization step where the orbitals of the hole and the excited electron are frozen has been presented and shown to be highly effective in converging on charge-localized solutions when applied in calculations of intramolecular charge transfer excited states of organic molecules with the LDA and GGA functionals. Significantly, in the case of the A$_2$ state of benzonitrile, the charge-localized solution could be obtained even if the electronic distance from the initial guess consisting of ground state canonical orbitals with non-aufbau occupation is larger than the distance to a charge-delocalized solution ($\eta$ of 0.49 and 0.19, respectively, when using the PBE functional). An extensive assessment of the robustness and efficiency of the method in comparison with other algorithms for orbital optimized excited state calculations, such as MOM, and the application to hybrid functionals will be presented in future works. 

The freeze-and-release DO strategy has enabled a benchmark study of the performance of orbital optimized density functional calculations with local and semi-local functionals for a large set of charge transfer excitations in organic molecules. Orbital optimized calculations with the LDA and the PBE and BLYP functionals are found to provide on average better estimates of the excitation energy compared to linear-response TD-DFT calculations with the same functionals. Most importantly, the TD-DFT errors tend to increase with the charge transfer distance. As a consequence, in cases where a molecule has excited states with different characters, TD-DFT can produce an incorrect order of the states because the energy of states with strong charge transfer is underestimated more than those with weak charge transfer, as exemplified by the B$_2$ and A$_1$ states of planar PP. In contrast, for the orbital optimized approach, excitations with strong charge transfer character are described with similar accuracy as excitations with a more local character. For the excitations with strong charge transfer, the accuracy of the orbital optimized calculations with the LDA, and the PBE and BLYP functionals rivals the accuracy of TD-DFT with more computationally demanding hybrid functionals\cite{Loos2021}. When comparing charge transfer distances, the deviation between the results of orbital optimized and TD-DFT calculations is small, even for the states with strong charge transfer character, where TD-DFT with semi-local functionals gives large errors on the excitation energy. However, a few limitations of the TD-DFT calculations should be noted. Firstly, even when the full, relaxed TD-DFT difference density matrix is used, the charge transfer distance obtained when using local and semi-local functionals tends to be large in comparison to results obtained in orbital optimized calculations. Therefore, linear-response TD-DFT seems to lack relaxation effects compared to the orbital optimized calculations, which are fully variational. Secondly, the charge transfer distance computed with TD-DFT can be too low in some cases because of mixing with Rydberg states when using diffuse functions, an issue that does not affect the orbital optimized calculations.

The balanced treatment of several electronic states using computationally efficient functionals combined with the robustness of DO can be especially useful for large molecular dynamics simulations in the condensed phase, where lowering the computational cost is crucial for obtaining statistically significant results, but previous orbital optimized approaches have suffered from convergence issues\cite{Mazzeo2023}. Despite these advantages, the absolute errors obtained here for local and semi-local functionals for the excitation energy can still be relatively large (up to $1.1$ eV), while many applications require a higher accuracy. Orbital optimized calculations of excited states can be affected by the delocalization error inherent in local and semi-local KS functionals, which can lead to the errors on the excitation energy observed here. The issue is evident in the case of the A$_1$ excited state of PP. Previous high-level calculations show minor changes in the character of the excitation with variations in molecular geometry\cite{Loos2021}. However, the solution obtained from orbital optimized calculations at the planar geometry has a relatively high degree of charge delocalization and no charge-localized solution closer to the initial guess has been found, compared to the solution obtained for the same state at the twisted geometry. The presence of charge-delocalized solutions is found to affect both the calculations of the spin-mixed and triplet excited states. Here, state specific methods accounting for the multi-determinant nature of open-shell singlet states in the orbital optimization, such as the restricted open-shell KS approach\cite{Kowalczyk2013, Frank1998}, might have an advantage over single-determinant methods, as they avoid calculating both a spin-mixed and a triplet solution.

Hybrid functionals can alleviate the delocalization error inherent in the KS approach. It is promising that for a selection of the excited states considered here, a significant improvement of the orbital optimized calculations is obtained when including exact exchange through the B3LYP and CAM-B3LYP functionals for both short- and long-range charge transfer, with the best results obtained for the range-separated functional CAM-B3LYP. For the three selected excited states with strong charge transfer character ($d^{\mr{CT}}>1.5$ Å), the orbital optimized calculations with the CAM-B3LYP functional provide errors around $-0.15$ eV. Here, TD-DFT using CAM-B3LYP (errors below 0.1 eV) outperforms the orbital optimized calculations. The good performance of TD-DFT with CAM-B3LYP for intramolecular charge transfer excitations is known\cite{Mester2022, Loos2021, Dev2012} and can be explained by the fact that this functional has in part been optimized to give accurate excitation energy values for this class of excitations\cite{Yanai2004}. However, it should be noted that TD-DFT 
calculations with CAM-B3LYP tend to overestimate the excitation energy of states with valence and short-range charge transfer excitation character, as found here for two states with $d^{\mr{CT}}<1.5$ Å and observed previously for similar systems\cite{Dev2012}. Moreover, TD-DFT with the CAM-B3LYP functional is known to fail for long-range intermolecular charge transfer excited states with errors of several eV\cite{Mester2022}. The orbital optimized calculations tend to overestimate the excitation energy of the short-range charge transfer excitations less and are expected to perform better for long-range intermolecular charge transfer excited states based on preliminary studies\cite{Hait2021, Barca2018}. A future task in order to confirm these advantages of the orbital optimized approach is to extend the benchmark of calculations with global and range-separated hybrid functionals to a larger set of intramolecular charge transfer excited states as well as intermolecular charge transfer states leveraging the robustness of DO approaches. It will also be interesting to assess the accuracy of the explicit inclusion of the self-interaction correction in calculations of charge transfer excitations, which has for example been shown to lead to a significant improvement of the excitation energy in calculations of a doubly excited state with charge transfer character in the ethylene molecule\cite{Schmerwitz2022, Levi2020b}.

\begin{acknowledgement}
This work was funded by the Icelandic Research Fund (grants nos. 217734, 217751, and 239678) and the Student Innovation Fund of the Icelandic Centre for Research. The calculations were carried out at the Icelandic High Performance Computing Center (IHPC) facility. The authors thank Aleksei V. Ivanov and Hannes Jónsson for useful discussions.

\end{acknowledgement}

\begin{suppinfo}


\end{suppinfo}

Convergence of the excitation energy for orbital optimized calculations with respect to the basis set; excitation energy from the orbital optimized and TD-DFT calculations with the LDA and the BLYP functional as a function of the charge transfer distance; charge transfer distances for the orbital optimized calculations with the LDA and the BLYP functionals.

\bibliography{main}

\end{document}


\begin{figure}[h!]
    \centering
    \includegraphics[width=0.6\textwidth]{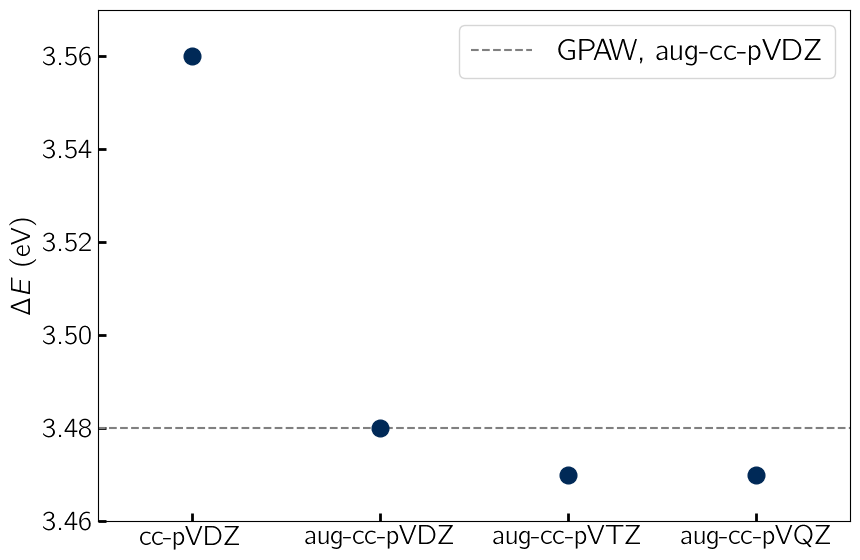}
    \caption{The excitation energy from all-electron orbital optimized calculations using the PBE functional with ORCA for the mixed-spin state for the B2 excitation of quinoxaline as a function of the basis set size. The corresponding excitation energy from GPAW, which uses the frozen core approximations, with the aug-cc-pVDZ+sz basis set is indicated with the dashed line. The good agreement between the ORCA and GPAW results show that in this case the effects of the frozen core approximation are not significant.}
   \label{fig:basis}
\end{figure}

\begin{figure}[h!]
    \centering
    \includegraphics[width=0.8\textwidth]{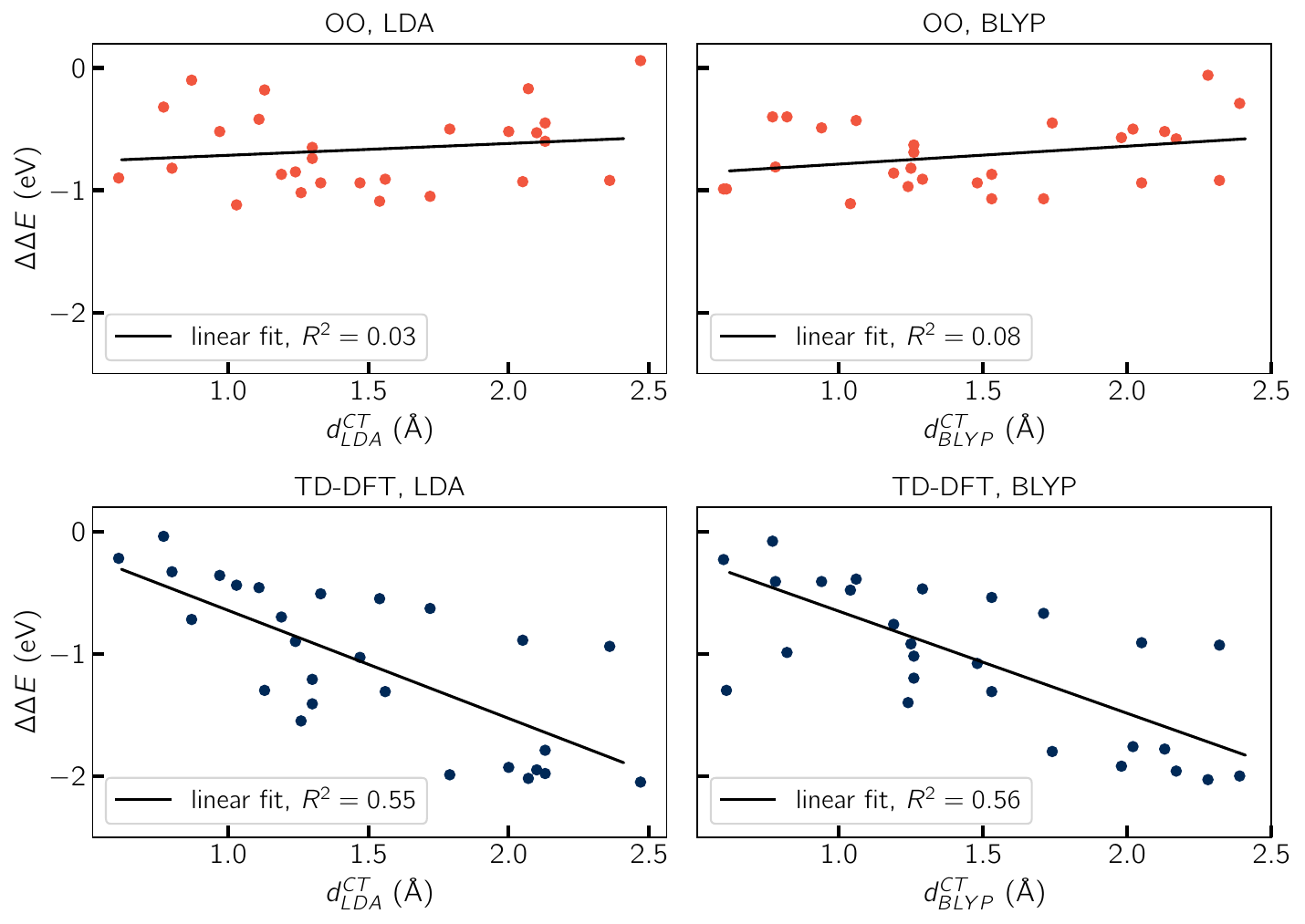}
    \caption{The error on the excitation energy relative to the theoretical best estimates from reference \citen{Loos2021} for the orbital optimized calculations with direct optimization and TD-DFT calculations using the LDA and the BLYP functional. For each functional, the error is shown as function of the charge transfer distance $d^{\mathrm{CT}}$. The black lines represent linear regression fits.}
   \label{fig:enscat_si}
\end{figure}

\begin{table}[h!]
  \caption{The charge transfer distance $d^{\mr{CT}}$ (in Å) obtained from orbital optimized calculations using the LDA and the BLYP functional.}
  \label{tbl:dist}
  \begin{tabular}{lc|ll}
    \hline
    
    Molecule  & Sym. &   \multicolumn{2}{c}{$d^{\mr{CT}}$}     \\
    \hline
     &  & LDA & BLYP   \\
    \hline
    Aminobenzonitrile (ABN)   & A$_1$ & 1.03 &   1.04  \\
    Aniline   & A$_1$ & 0.80 &  0.78       \\
    Azulene   &  A$_1$ & 0.97    & 0.94  \\
       & B$_2$ & 0.77 &  0.77   \\
    Benzonitrile   & A$_2$  & 0.47 & 0.61     \\
    Benzothiadiazole (BTD) & B$_2$  & 1.19 & 1.19    \\
    Dimethylaminobenzonitrile (DMABN) & A$_1$  &1.54   & 1.53  \\ 
    DMABN, twisted &  A$_2$ & 2.10  & 2.02   \\
    & B$_1$  & 1.79 &  1.74    \\
    Dimethylaniline (DMAn) & B$_2$  & 1.11  & 1.06    \\
    &  A$_1$ & 1.33 &  1.29   \\
    Hydrogen Chloride (HCl) & $\Pi$  & 0.87 & 0.82        \\
    Nitroaniline & A$_1$  &2.05  & 2.05   \\
    Nitrobenzene & A$_1$  & 1.47 &  1.48   \\
    Nitrodimethylaniline (NDMA) & A$_1$  & 2.36 & 2.32   \\
    Nitropyridine N-Oxide (NPNO) & A$_1$  & 1.72 &  1.71   \\
    $N$-Phenylpyrrole (PP) & B$_2$  & 1.56 & 1.53   \\
    &  A$_1$ & 2.00 & 1.98    \\
    PP, twisted & B$_2$ & 2.07 & 2.39   \\
    &  A$_1$ & 2.47 & 2.28    \\
    & A$_2$  & 2.13 & 2.13      \\
    & B$_1$  & 2.13 &  2.17    \\
    Phthalazine &  A$_2$ & 1.30 & 1.26    \\
    &  B$_1$ & 1.30 &  1.26     \\
    Quinoxaline & B$_2$  & 1.24 & 1.25   \\
    & A$_1$  & 0.61 &  0.60     \\
    & B$_1$  & 1.26 &  1.24  \\

    \hline
  \end{tabular}
\end{table}

\clearpage

\bibliography{si}